\documentclass[twocolumn,preprintnumbers,amsmath,amssymb,floatfix,prl,showpacs]{revtex4}
\usepackage{graphicx}% Include figure files
\usepackage{bm}% bold math

\usepackage{graphics}
\usepackage{color}

\begin{document}

%\preprint{APS/123-QED}

\title{Theory of the ac spin-valve effect}
% Force line breaks with \\

\author{Denis Kochan}
\author{Martin Gmitra}
\author{Jaroslav Fabian}
%\email{kochan@fmph.uniba.sk}
\affiliation{Institute for Theoretical Physics, University of Regensburg, 93040 Regensburg, Germany}

%\date{\today}

\begin{abstract}
\noindent The spin-valve complex magnetoimpedance of symmetric ferromagnet/normal
metal/ferromagnet junctions is investigated within the drift-diffusion (standard) model of spin injection.
The ac magnetoresistance---the real part difference of the impedances of
the parallel and antiparallel magnetization configurations---exhibits
an overall damped oscillatory behavior, as an interplay
of the diffusion and spin relaxation times. In wide junctions the
ac magnetoresistance oscillates between positive and \emph{negative} values, reflecting
resonant amplification and depletion of the spin accumulation, while the line shape
for thin tunnel junctions is predicted to be purely Lorentzian. The ac spin-valve effect
could be a technique to extract spin transport and spin relaxation parameters in
the absence of a magnetic field and for a fixed sample size.
\end{abstract}

%\pacs{XX.xx.XX}
 \maketitle

Electrical spin injection from a ferromagnetic ($F$) to a nonmagnetic ($N$)
conductor is essential for spintronics \cite{Zutic2004:RMP, Fabian2007:ActaPhysSlov}.
Predicted by Aronov \cite{Aronov1976:ZETF} and first realized by Johnson and Silsbee,
\cite{Johnson1985:PRL,Johnson1987:PRB,Johnson1988:PRB}, it
is now a well established concept. A biased ferromagnetic/nonmagnetic junction
generates a nonequilibrium spin accumulation within the spin diffusion length
at the interface, building a nonequilibrium resistance
\cite{Johnson1987:PRB,vanSon1987:PRL,Rashba2002:EurPhysJB}.
In an $FNF$ junction this nonequilibrium resistance gives rise to the difference
in the junction resistances for parallel (P) and antiparallel (AP) magnetization orientations of the $F$
regions---the giant magnetoresistance (GMR) \cite{Baibich1988:PRL,Binasch1989:PRB}.
Drift-diffusion theory along with the spin accumulation concept successfully describes
magnetoresistance effects in charge neutral \cite{Valet1993:PRB,Hershfield1997:PRB}
as well as in space-charge systems \cite{Zutic2001:PRB,Fabian2002:PRB},
enabling one to obtain relevant spin-related materials parameters \cite{Bass2007:JPCM},
such as the spin relaxation times.

Recently, Rashba has generalized the spin-polarized drift-diffusion theory to the alternating current
(ac) regime \cite{Rashba2002:APL,Rashba2002:EurPhysJB}.
We apply this theory and investigate the complex
impedance ${\cal Z}(\omega)$ of symmetric $FNF$ junctions.
We show that the real part of the spin-valve magnetoimpedance (we
call it here ac magnetoresistance) $\varDelta\mathcal{Z}=\mathcal{Z}_{\rm AP}-\mathcal{Z}_{\rm P}$
of the junctions exhibits damped oscillations as a function of frequency. The oscillation period
is given by the diffusion time through the normal layer. In mesoscopic junctions
(of sizes up to the spin relaxation length $L_s$), the ac magnetoresistance
can be {\it negative} at experimentally accessible frequencies,
meaning that the antiparallel configuration has a lower ac resistance
than the parallel one. The negative ac magnetoresistance is a consequence
of a resonant spin accumulation effect, namely a resonant spin amplification in the P configuration
and a resonant spin depletion in the AP one. In nanoscale junctions (with sizes
much less than $L_s$), with tunnel
contacts, the oscillation period is large, leaving a nice Lorentzian profile
with the width of the spin relaxation rate.
A one-parameter fit to the line shape (either damped oscillator or Lorentzian)
determines the spin relaxation time $\tau_s$.

We present the ac spin-valve effect as an alternative to other
methods that measure $\tau_s$ of nonmagnetic conductors, such as the conduction electron
spin resonance, spin pumping, or the Hanle effect, which require magnetic fields, or to the
dc spin injection method (in vertical or lateral geometries), which requires
studying various sample sizes (distances to electrodes) to extract
the spin diffusion length \cite{Jedema2001:Nature}. In a sense the
ac spin-valve effect is similar to the
Hanle effect, which is widely used to find spin relaxation times in metals
and semiconductors \cite{Johnson1985:PRL, Huang2007:PRL},
but the role of the magnetic field is taken by the frequency;
in the Hanle effect too the signal in general oscillates as a function of magnetic
field, with a modified Lorentzian shape in the diffusive regime \cite{Fabian2007:ActaPhysSlov}.

\begin{figure}[h!]
 \includegraphics[angle=0,width=0.5\columnwidth]{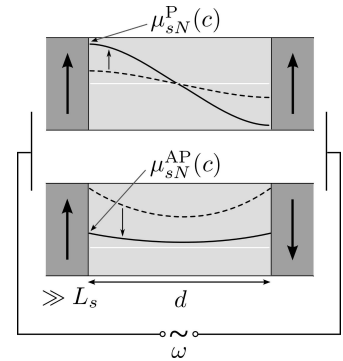}
 \caption{Scheme of an $FNF$ spin valve. The spacer $N$ region
 has width $d$ and the sizes of the ferromagnetic electrodes
 are assumed greater than the corresponding spin diffusion length
 $L_s$. In the dc regime the parallel configurations results
 in smaller spin accumulation (dashed line) than in the antiparallel one,
 demonstrated by the positive dc spin-valve magnetoresistance.
 In the ac regime, this can be reversed (solid): at certain frequency
 ranges there can be a resonant spin amplification in the parallel and spin depletion in the
 antiparallel configuration, resulting in a negative
 ac magnetoresistance.
 }\label{Fig:scheme}
\end{figure}

A microwave measurement of $\tau_s$ in the absence of
a magnetic field may be important as in many conductors
the spin relaxation time depends strongly on it;
a striking case is aluminum in which $\tau_s$ decreases by an order of magnitude
as the magnetic field increases from 0.05 to 1.3 T \cite{Lubzens1976:PRL}.
Still, $\tau_s$ obtained by spin resonance tend to be, for a given temperature, much
greater than that obtained from transport techniques, as catalogued for Al and Cu
in Ref. \onlinecite{Jedema2003:PRB}. The case of Au is even more striking,
as spin resonance shows that at low temperatures the ratio of $\tau_s$ to the momentum
relaxation time is about one, while transport techniques predict the ratio to be
about 100 \cite{Monod1977:JLTP, Johnson1993:PRL}; at room temperature, at which phonons
are relevant, the ratio is about 10, as measured by spin pumping \cite{Mosendz2009:PRB}
which requires both magnetic field and nanoscale transparent junctions.
For extracting bulk spin relaxation times it may be preferable to work
with tunnel contacts and mesoscopic samples, so that
spin relaxation is not strongly influenced by the interface and surface effects.
(Various techniques for measuring $\tau_s$ as well as useful data are given in
the review Ref. \onlinecite{Bass2007:JPCM}.) The ac spin-valve method
could potentially explore nano and mesoscopic spin valves, in both vertical and
lateral geometries, at no magnetic field applied to the normal conductor,
and provide the spin relaxation times at a fixed sample size
\endnote{Spurious effects are eliminated since one takes the resistance difference for
the parallel and antiparallel magnetizations of the
ferromagnetic layers. The skin effect at the relevant frequencies will
not be an issue, if the junction sizes are micrometers or less, even for perfect
conductors such as Cu. As there is no external magnetic field, transmission spin resonance/spin pumping
effects will also not be present.}.

%Before proceeding further let us recall two well
%known properties of the spin valve resistance
%$\varDelta\mathcal{R}$ in collinear geometry and dc regime,
%$\varDelta\mathcal{R}=\varDelta{\cal Z}(\omega=0)$. Firstly,
%$\varDelta\mathcal{R}>0$, i.e. the resistance of the $F/N/F$
%junction in the AP configuration prevails its P configuration
%counterpart. This phenomenon is known as the positive GMR.
%Secondly, the spin valve effect becomes significantly suppressed
%when thickness of the $N$ is much larger then its spin diffusion
%length, $d\gg L_{sN}$, in this case $\varDelta\mathcal{R}\approx
%e^{-d/L_{sN}}\approx 0$. The optimal experimental resolution for
%detecting the spin valve effect appears when $d\approx L_{sN}$.

%-------------

We consider a symmetric $FNF$ junction as comprising two $FN$ junctions in series, see
Fig.~\ref{Fig:scheme}.
Each $FN$ junction has a contact ($c$) region with a spin-dependent
conductance; otherwise spin is assumed to be preserved at the contact.
The spin valve dc magnetoresistance $\varDelta\mathcal{R} = \varDelta{\cal Z}(\omega=0)$
of a symmetric $FNF$ junction, whose $N$ region has width $d$ and the $F$ regions
have widths much greater than the spin diffusion lengths,
can be expressed analytically within the drift-diffusive regime
\cite{Rashba2002:APL,Rashba2002:EurPhysJB, Zutic2004:RMP, Fabian2009:Julich}.
This dc formula has a straightforward extension to the harmonic ac regime,
and we write the complex magnetoimpedance as
\begin{align}\label{spin-valve impedance}
&\varDelta\mathcal{Z}(\omega,d)=\\
&\dfrac{8\, r_N(\omega)[r_F(\omega)\, P_{\sigma_F}+r_c\, P_{\,\Sigma_c}]^2\,e^{d/L_{sN}
(\omega)}}{[r_F(\omega)+r_c+r_N(\omega)]^2\,e^{2d/L_{sN}(\omega)}-[r_F(\omega)+r_c-r_N(\omega)]^2},
\nonumber
\end{align}
by indicating the complex frequency-dependent quantities (labeled by the region $N$ and $F$),
\begin{align}
\tau_{s{}}(\omega) &= \tau_{s{}}\bigl/\bigl(1-i\omega\tau_{s{}}\bigr),\label{transformation1}\\
L_{s{}}(\omega) &= L_{s{}}\bigl/\sqrt{\smash[b]{1-i\omega\tau_{s{}}}}, \label{transformation2}\\
r_{{}}(\omega) &= r_{{}}\bigl/\sqrt{\smash[b]{1-i\omega\tau_{s{}}}}. \label{transformation3}
\end{align}
Here $\tau_{s}$ is the spin relaxation time, $L_s=\sqrt{\smash[b]{D\tau_s}}$ is the spin diffusion length,
$D$ is the diffusivity, and $r = L_s/\sigma$ is the effective resistance, with $\sigma$ denoting
the conductivity. The effective contact resistance
is $r_c = (\Sigma_\uparrow + \Sigma_\downarrow)/4\Sigma_\uparrow\Sigma_\downarrow$, with
$\Sigma_\lambda$ the contact conductance of spin $\lambda$.
Finally, $P_{\sigma_F}$ and $P_{\,\Sigma_c}$  denote the spin polarization of
the conductivity and conductance of the $F$ and contact region, respectively.
Driving ac is assumed to be harmonic with the angular frequency
$\omega=2 \pi f$, i.e. $j(t)\propto e^{-i\omega t}$.

We analyze the spin-valve impedance, based on Eq.~(\ref{spin-valve impedance}), for
a realistic model Py/Cu/Py junction, with the following experimentally obtained
data \cite{Jedema2001:Nature,Jedema2003:PRB} at the temperature $T=4.2\,{\rm K}$:
$L_{sN}=1\,{\rm \mu m}$, $\tau_{sN}=42\,{\rm{ps}}$, $D_N=238\,{\rm cm^2\,s^{-1}}$, $r_N=14\,{\rm f\Omega\,m^{2}}$,
$L_{sF}=5.5\,{\rm nm}$, $\tau_{sF}=0.6\,{\rm{ps}}$, $D_F=0.5\,{\rm cm^2\,s^{-1}}$, $r_F=0.42\,{\rm f\Omega\,m^{2}}$,
$P_{\sigma_F}=0.22$. For the contact characteristics we employ \cite{Takahashi2003:PRB}:
$r_c(\simeq r_F)=0.5\,{\rm f\Omega\,m^{2}}$ and $P_{\,\Sigma_c}=0.4$, so the contact interface
is generic, neither tunnel nor transparent.
The specific spin resistivities $r_F$,
$r_N$ and $r_c$ and hence the spin valve $\varDelta\mathcal{Z}(\omega,d)$
are evaluated for a unit cross section. In the experiment
one divides these resistivities by the actual conductor
cross sections, which could be $10^{-3}-1\,{\rm \mu m^{2}}$.

\begin{figure}[h!]
 \includegraphics[width=1.0\columnwidth]{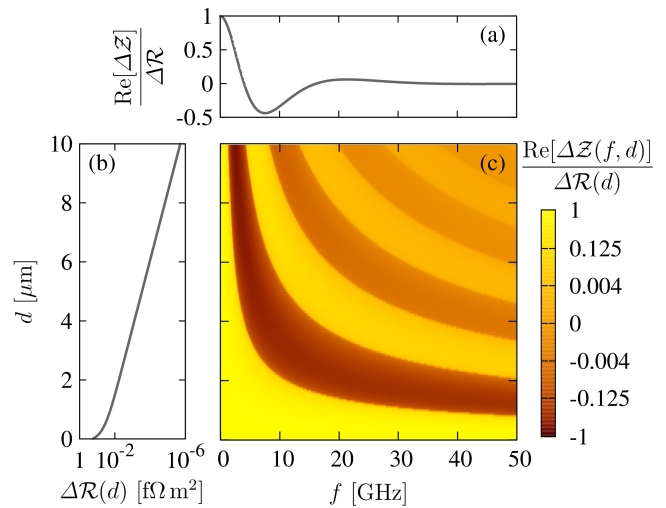}
 \caption{(color online) ac spin-valve effect in a model Py/Cu/Py junction.
(a)~Calculated ac/dc ratio of the spin-valve magnetoresistance as a function of $f=\omega/2\pi$
for $d=4\,{\rm \mu m}$.
(b)~Calculated dc spin-valve magnetoresistance as a function
of $d$.
(c)~Calculated ac/dc ratio of the spin-valve magnetoresistance as a function of $d$
and driving frequency $f$. The visible light and dark bands of equal signs are
separated by the node lines, $\mathrm{Re}[\varDelta\mathcal{Z}(f,d)]=0$.
 }\label{Fig:map}
\end{figure}

Figure~\ref{Fig:map} presents the calculated magnetoresistance.
In Fig.~\ref{Fig:map}(b) we show the dc magnetoresistance as a function of the $d$.
With increasing $d$ the magnetoresistance exponentially decreases,
as the injected spin accumulation is damped.
The plot in Fig.~\ref{Fig:map}(c) shows the ratio of the ac to the dc magnetoresistance,
$\mathrm{Re}[\varDelta\mathcal{Z}(f,d)]\bigl/\varDelta\mathcal{R}(d)$,
as a function of $d$ and frequency $f=\omega/2\pi$. For a given $d$, the ac magnetoresistance
oscillates as a function of $f$, between positive and negative values.
In Fig.~\ref{Fig:map}(a) the oscillations are shown for $d=4\,{\rm \mu m}$.
The negative peaks are considerable fractions (tens of percents) of the dc values.
On the $\mbox{\emph{d-f}}$-plot the oscillations show hyperbolic stripes. For thin samples,
the dependence on $f$ is rather weak for this generic junction. We will see
below that for tunnel junctions the dependence becomes Lorentzian.

%------------------------------------------------------------------------
%------------------------------------------------------------------------

To be specific, consider $d=L_{sN}=1\,{\rm \mu m}$.
The ac magnetoresistance remains positive for $f < 34.8\,{\rm GHz}$. Further
increase in the driving frequency leads to a negative ac spin-valve
magnetoresistance: $\mathrm{Re}(\varDelta\mathcal{Z}) < 0$. For $d=3L_{sN}=3\,{\rm \mu m}$ the
spin-valve magnetoresistance remains positive up to the frequency
$f\approx 6\,{\rm GHz}$, then it becomes negative
for $6\,{\rm GHz}\lesssim f\lesssim 26.1\,{\rm GHz}$.
There should be more oscillations observable at larger values of $d$,
but at the cost of exponentially reducing the magnitude, see  Fig.~\ref{Fig:map}(a).
We will see below that the relevant time scale parameter for the
oscillations is the diffusion time through the spacer layer.
For our model junction, a reasonable parameter
range for measuring the ac oscillations would be the sample sizes $L_{sN} \lesssim d \lesssim 4L_{sN}$.
The involved frequency,
$f = \omega/2\pi$, ranges are $1\,\mathrm{GHz}-50\,\mathrm{GHz}$,
experimentally well accessible.

Mathematically, the spin valve oscillations appear naturally. The real dc transport
parameters become in the ac case complex, see
Eqs.~(\ref{transformation1})-(\ref{transformation3}). The imaginary part of
$d/L_{sN}(\omega)$ gives rise to the ac exponential
$e^{d/L_{sN}(\omega)}$ in Eq.~(\ref{spin-valve impedance}) with
the trigonometric character and hence a certain oscillatory
behavior of the complex spin valve impedance $\varDelta\mathcal{Z}(\omega)$.
For the frequencies $\omega\ll\tau_{sN}^{-1}(\ll\tau_{sF}^{-1})$
the imaginary part of $d/L_{sN}(\omega)$
plays no role, see Eq.~(\ref{transformation2}). The
ac magnetoresistance exhibits changes on the scales of the
relaxation rate $1/\tau_s$ or the diffusion rate through the
spacer. These provide the practical limit for the use
of microwaves in the experiment.

%With increasing $d$ the spin valve oscillations are observable
%at low frequencies but their magnitude is exponentially suppressed.
%On the other hand the larger $d$ the lower frequencies are needed to drive
%junction to the negative magnetoimpedance regime.

We now give a qualitative picture of the predicted oscillatory behavior,
including the negative ac spin-valve magnetoresistance. First, we show
that the spin valve impedance
$\varDelta\mathcal{Z}(\omega)$ is related to the contact values of
the spin accumulations in $N$, for P and AP configurations.
From the standard spin injection model for a symmetric $FNF$
junction we derive the following formula \cite{unpublished}:
\begin{equation}\label{cornerstone}
\dfrac{\mu_{sN}^{\mathrm{P}}(c,t)-\mu_{sN}^{\mathrm{AP}}(c,t)}{j(t)}=
\dfrac{r_F(\omega)+r_c}{r_F(\omega)P_{\sigma_F}+r_cP_{\,\Sigma_c}}\,
\varDelta\mathcal{Z}(\omega)\,.
\end{equation}
Here $\mu_{sN}^{\mathrm{P/AP}}(c,t)$ represent the actual nonequilibrium spin
accumulation in the $N$ spacer for P and AP configurations
respectively, at the left $FN$ contact interface $c$ (see Fig.~\ref{Fig:scheme})
and $j(t)$ is the driving harmonic ac.
To understand the ac magnetoresistance oscillations, one
needs to look at the contact spin accumulation only.

%\textcolor{red}{Consider an ac-driven $FNF$ junction and let us
%investigate the spin packets(spin probability density)
%injected/extracted at $t=0$ and later times.

\begin{figure}[h!]
 \includegraphics[width=0.99\columnwidth]{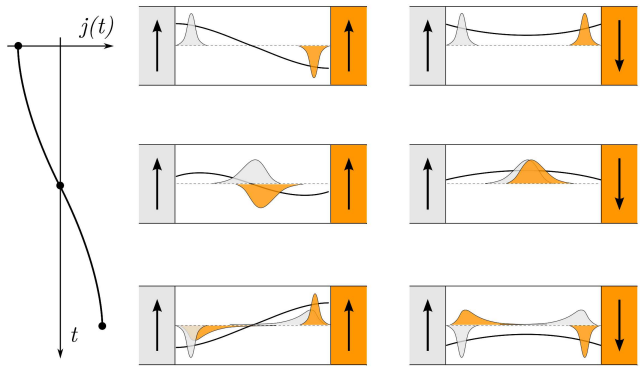}
 \caption{(color online) Mechanism for the resonant amplification and depletion of the spin accumulation
          in an ac-driven $FNF$ junction for parallel (P) and antiparallel (AP) configurations.
          The solid lines within the horizontal of the $N$ spacer represent actual
          profiles of the nonequilibrium
          spin accumulation $\mu_{sN}^{\mathrm{P/AP}}(x,t)$, which correspond to the harmonic ac signal $j(t)$ shown on
          the left.
          The injected and extracted spin packets and their diffused and spread positions are shown at three distinct
          times $t=0$, $t=T_N/4$, and $t=T_N/2$, where $T_N$ is the characteristic diffusion time across the $N$ conductor.
          Frequency of the driving ac is close to $1/T_N$ and tones of the packets correspond to $F$ conductors
          which initially emitted them.
 }\label{Fig:toy-model}
\end{figure}

The qualitative picture is in Fig.~\ref{Fig:toy-model}, which
shows P and AP configurations at three times $t=0$, $t=T_N/4$, and $t=T_N/2$,
where $T_N=\tau_{sN}\,d^{{2}}/L_{sN}^2=d^2/D_N$ is the diffusion time
through $N$. The resonant spin amplification and depletion effect happens
if the driving current $j(t)$ is close to the $N$ spacer diffusion time $T_N$,
this case is shown in Fig.~\ref{Fig:toy-model}.

%When supposing $d=L_{sN}$ the characteristic spin diffusion time equals
%relaxation time and injected spin packets at one interface arrives to the
%other one with decreased amplitude $1/e$.
%------------------------------------------------------------------------
%------------------------------------------------------------------------

At time $t=0$ the current $j$ is negative and electrons
are injected from the left and extracted to the right electrodes,
leaving behind positive and negative spin accumulations, indicated
in Fig.~\ref{Fig:toy-model} by diffusive packets
(the sample is locally charge neutral, only spin is redistributed
nonuniformly). The dynamics of these spin packets is governed by diffusion
and relaxation, but by not bias voltage. This is because in the $N$ spacer, there is no
spin-charge coupling and spin and charge transports are decoupled, see
[\onlinecite{Fabian2007:ActaPhysSlov}].
At time $t=T_N/4$ the current vanishes, $j=0$, as
well as the spin injection and extraction. In the meantime
the spin packets diffusively spread and reach the center of the $N$ spacer.
At $t=T_N/2$ the spin packets reach the other contacts. Now
the current is fully reversed: in the P configuration
the new spin packet is injected at the right electrode, {\it amplifying}
the initial injected spin packet that has traveled from the left.
In the AP configuration, the new spin packet of the opposite
sign is injected at the right, {\it depleting} the initial
injected spin. Similarly at the left electrode.
The left contact difference
$\mu_{sN}^{\mathrm{P}}(c)-\mu_{sN}^{\mathrm{AP}}(c)$ at $t=T_N/2$ becomes
negative, the actual current $j>0$ and, according to Eq.~(\ref{cornerstone}), we get
negative ac magnetoresistance, $\mathrm{Re}(\varDelta\mathcal{Z})<0$.

The resonance condition, $\omega T_N \approx\pi$, is equivalent to
$L_{sN}/d \approx \sqrt{\omega \tau_{sN}/\pi}$. In practice, to see the
negative ac magnetoresistance one prefers $d \approx L_{sN}$, so that
$\omega \tau_{sN} \approx\pi$, which is the microwave regime. If
$d \agt L_{sN}$, as in our model shown in Fig.~\ref{Fig:map}, then
the oscillations can be observed at lower frequencies, but at the
cost of decreasing the magnitude of the ac magnetoresistance due
to spin relaxation. This  need not be an issue with tunnel contacts,
as the precision of measuring higher resistances is higher. On the
other hand, no oscillations (within the GHz regime) should be seen
for nanoscale junctions, for $d \ll L_{sN}$. We will show below
that this important regime gives a Lorentzian profile.

\begin{figure}[h!]
 \includegraphics[width=0.85\columnwidth]{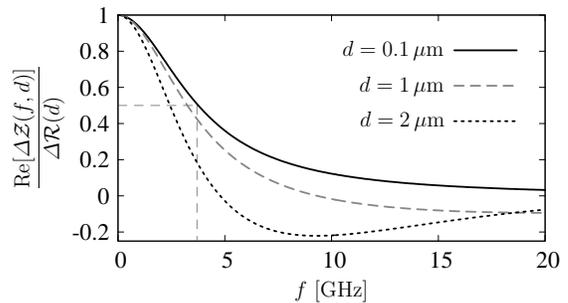}
 \caption{Calculated ac/dc ratio of the $FNF$ magnetoresistances
          for three different widths $d$ of the $N$ spacer with tunnel contacts to $F$.
          The solid line, $d=0.1\,{\rm \mu m}$, represents a Lorentzian line shape with its half-width
          determining the spin relaxation time $\tau_{sN}$. The dotted line, $d=2\,{\rm \mu m}$,
          shows the first oscillation, with negative ac magnetoresistance. The dashed line, $d=1\,{\rm \mu m}$,
          is the intermediate case. The values for the parameters
          are as in Fig.~\ref{Fig:map}, but with a greater tunnel resistance, $r_c=5\,{\rm n\Omega\, m^2}$.
         }
 \label{Fig:4}
\end{figure}

We now turn to the case of a junction with tunnel contacts, $r_F,r_N\ll r_c$.
(In general, using tunnel barriers allows to adopt the junction resistance and
size independently, providing maximal flexibility in the device design.)
In this case Eq.~(\ref{spin-valve impedance}) reduces to
\begin{equation}\label{tunnel spin-valve impedance1}
\varDelta\mathcal{Z}
\approx\dfrac{4\, r_N}{\sqrt{\smash[b]{1-i\omega\tau_{sN}}}}\dfrac{P_{\,\Sigma_c}^2}{\sinh{\bigl[\tfrac{d}{L_{sN}}\sqrt{\smash[b]{1-i\omega\tau_{sN}}}\bigr]}}\,,
\end{equation}
where $L_{sN}=\sqrt{\smash[b]{D_N\tau_{sN}}}$.
A single parameter $(\tau_{sN}$, knowing $d$ and $D_N$) fit of a measurement of the
$\omega$ dependence of the tunnel spin valve impedance (relative to the dc value) to  Eq.~(\ref{tunnel spin-valve impedance1})
can determine the spin relaxation time of the normal region. The shape is illustrated in Fig. \ref{Fig:4}.
Since $\tau_{sF}$ is typically an order or two magnitudes
smaller than $\tau_{sN}$, the ac effects do not play significant in the $F$ electrodes.

For $d \gtrsim |L_{sN}(\omega)|$ we can approximate $\varDelta\mathcal{Z}$ as follows:
\begin{equation}\label{tunnel spin-valve impedance3}
\varDelta\mathcal{Z}(\omega,d,\tau_{sN})
\approx 8\dfrac{r_N\,P_{\,\Sigma_c}^2}{\sqrt{\smash[b]{1-i\omega\tau_{sN}}}}\,e^{-d\sqrt{\tfrac{1-i\omega\tau_{sN}}{D_N\tau_{sN}}}}\,.
\end{equation}
Suppose we know the experimental value of the frequency
$\omega_0$ at which ${\rm Re}[\varDelta\mathcal{Z}(\omega_0,d)]$ vanishes. As an
alternative to the fitting, the spin relaxation time can be given by the equation:
\begin{equation}
\dfrac{1+\sqrt{\smash[b]{1+\omega_0^2\tau_{sN}^2}}}{\omega_0\tau_{sN}}=
\tan{\dfrac{d\,\omega_0}{\sqrt{2D_N}}\sqrt{\dfrac{\tau_{sN}}{1+\sqrt{\smash[b]{1+\omega_0^2\tau_{sN}^2}}}}}\,,
\end{equation}
which can be solved for $\tau_{sN}$ with simple numerics.

In the opposite important case of $d\ll L_{sN}$, Eq. (\ref{tunnel spin-valve impedance1}) becomes a Lorentzian:
\begin{equation}\label{tunnel spin-valve impedance2}
\varDelta\mathcal{Z}(\omega,d,\tau_{sN})\approx 4\dfrac{r_N\,P_{\,\Sigma_c}^2}{d/\sqrt{\smash[b]{D_N\tau_{sN}}}}\,\dfrac{1+i\omega\tau_{sN}}{1+\omega^2\tau_{sN}^2}\,.
\end{equation}
The half-width frequency $\omega_{1/2}$ at which ${\rm Re}[\varDelta\mathcal{Z}(\omega_{1/2},d)]=\tfrac{1}{2}\varDelta\mathcal{R}(d)$
determines the spin relaxation time according to $\tau_{sN}=1/\omega_{1/2}$.
This Lorentzian shape is rather robust for tunnel junctions, illustrated in Fig~\ref{Fig:4},
which also shows an intermediate case of $d \approx L_{sN}$.

%------------------------------------------------------------------------
%------------------------------------------------------------------------

For transparent contacts ($r_c \ll r_F,r_N$) and nanoscale junctions, $d\ll L_{sN}$,
the magnetoimpedance is $\varDelta\mathcal{Z}(\omega,d)\approx 2 r_F(\omega) P^2_{\sigma_F}$,
the square root of the Lorentzian, but with the width of $1/\tau_{sF}$. The shape is therefore featureless
in the microwave regime, unless the ferromagnetic contacts have a relatively large spin relaxation
time, in which case $\tau_{sF}$ could be determined from the measured shape. In mesoscopic
junctions oscillations should be visible.

In summary, we have presented a simple but robust theory of the
ac spin-valve effect in symmetric $FNF$ junctions
and predict negative ac magnetoresistance due to resonant amplification and depletion of
the spin accumulation in the normal metal region. The oscillating line shape allows a single-parameter
(spin relaxation time) fitting for mesoscopic and nanoscale spin valves;
for the latter a Lorentzian shape should be seen with tunnel contacts.

We thank G.~Woltersdorf, C.~Back, and S.~Parkin for useful
discussions about possible experimental realizations of the ac spin-valve effect.
The work has been supported by the DFG SFB 689.

\bibliography{spin}

\begin{thebibliography}{26}
\expandafter\ifx\csname natexlab\endcsname\relax\def\natexlab#1{#1}\fi
\expandafter\ifx\csname bibnamefont\endcsname\relax
  \def\bibnamefont#1{#1}\fi
\expandafter\ifx\csname bibfnamefont\endcsname\relax
  \def\bibfnamefont#1{#1}\fi
\expandafter\ifx\csname citenamefont\endcsname\relax
  \def\citenamefont#1{#1}\fi
\expandafter\ifx\csname url\endcsname\relax
  \def\url#1{\texttt{#1}}\fi
\expandafter\ifx\csname urlprefix\endcsname\relax\def\urlprefix{URL }\fi
\providecommand{\bibinfo}[2]{#2}
\providecommand{\eprint}[2][]{\url{#2}}

\bibitem[{\citenamefont{\v{Z}uti\'{c} et~al.}(2004)\citenamefont{\v{Z}uti\'{c},
  Fabian, and Sarma}}]{Zutic2004:RMP}
\bibinfo{author}{\bibfnamefont{I.}~\bibnamefont{\v{Z}uti\'{c}}},
  \bibinfo{author}{\bibfnamefont{J.}~\bibnamefont{Fabian}}, \bibnamefont{and}
  \bibinfo{author}{\bibfnamefont{S.~D.} \bibnamefont{Sarma}},
  \bibinfo{journal}{Rev. Mod. Phys.} \textbf{\bibinfo{volume}{76}},
  \bibinfo{pages}{323} (\bibinfo{year}{2004}).

\bibitem[{\citenamefont{Fabian et~al.}(2007)\citenamefont{Fabian,
  Matos-Abiague, Ertler, Stano, and \v{Z}uti\'{c}}}]{Fabian2007:ActaPhysSlov}
\bibinfo{author}{\bibfnamefont{J.}~\bibnamefont{Fabian}},
  \bibinfo{author}{\bibfnamefont{A.}~\bibnamefont{Matos-Abiague}},
  \bibinfo{author}{\bibfnamefont{C.}~\bibnamefont{Ertler}},
  \bibinfo{author}{\bibfnamefont{P.}~\bibnamefont{Stano}}, \bibnamefont{and}
  \bibinfo{author}{\bibfnamefont{I.}~\bibnamefont{\v{Z}uti\'{c}}},
  \bibinfo{journal}{Acta Phys. Slov.} \textbf{\bibinfo{volume}{57}},
  \bibinfo{pages}{565} (\bibinfo{year}{2007}).

\bibitem[{\citenamefont{Aronov}(1976)}]{Aronov1976:ZETF}
\bibinfo{author}{\bibfnamefont{A.~G.} \bibnamefont{Aronov}},
  \bibinfo{journal}{JETP Lett.} \textbf{\bibinfo{volume}{24}},
  \bibinfo{pages}{32} (\bibinfo{year}{1976}).

\bibitem[{\citenamefont{Johnson and Silsbee}(1985)}]{Johnson1985:PRL}
\bibinfo{author}{\bibfnamefont{M.}~\bibnamefont{Johnson}} \bibnamefont{and}
  \bibinfo{author}{\bibfnamefont{R.~H.} \bibnamefont{Silsbee}},
  \bibinfo{journal}{Phys. Rev. Lett.} \textbf{\bibinfo{volume}{55}},
  \bibinfo{pages}{1790} (\bibinfo{year}{1985}).

\bibitem[{\citenamefont{Johnson and Silsbee}(1987)}]{Johnson1987:PRB}
\bibinfo{author}{\bibfnamefont{M.}~\bibnamefont{Johnson}} \bibnamefont{and}
  \bibinfo{author}{\bibfnamefont{R.~H.} \bibnamefont{Silsbee}},
  \bibinfo{journal}{Phys. Rev. B} \textbf{\bibinfo{volume}{35}},
  \bibinfo{pages}{4959} (\bibinfo{year}{1987}).

\bibitem[{\citenamefont{Johnson and Silsbee}(1988)}]{Johnson1988:PRB}
\bibinfo{author}{\bibfnamefont{M.}~\bibnamefont{Johnson}} \bibnamefont{and}
  \bibinfo{author}{\bibfnamefont{R.~H.} \bibnamefont{Silsbee}},
  \bibinfo{journal}{Phys. Rev. B} \textbf{\bibinfo{volume}{37}},
  \bibinfo{pages}{5312} (\bibinfo{year}{1988}).

\bibitem[{\citenamefont{van Son et~al.}(1987)\citenamefont{van Son, van Kempen,
  and Wyder}}]{vanSon1987:PRL}
\bibinfo{author}{\bibfnamefont{P.~C.} \bibnamefont{van Son}},
  \bibinfo{author}{\bibfnamefont{H.}~\bibnamefont{van Kempen}},
  \bibnamefont{and} \bibinfo{author}{\bibfnamefont{P.}~\bibnamefont{Wyder}},
  \bibinfo{journal}{Phys. Rev. Lett.} \textbf{\bibinfo{volume}{58}},
  \bibinfo{pages}{2271} (\bibinfo{year}{1987}).

\bibitem[{\citenamefont{Rashba}(2002{\natexlab{a}})}]{Rashba2002:EurPhysJB}
\bibinfo{author}{\bibfnamefont{E.~I.} \bibnamefont{Rashba}},
  \bibinfo{journal}{Eur. Phys. J. B} \textbf{\bibinfo{volume}{29}},
  \bibinfo{pages}{513} (\bibinfo{year}{2002}{\natexlab{a}}).

\bibitem[{\citenamefont{Baibich et~al.}(1988)\citenamefont{Baibich, Broto,
  Fert, NguyenVanDau, Petroff, Eitenne, Creuzet, Friederich, and
  Chazelas}}]{Baibich1988:PRL}
\bibinfo{author}{\bibfnamefont{M.~N.} \bibnamefont{Baibich}},
  \bibinfo{author}{\bibfnamefont{J.~M.} \bibnamefont{Broto}},
  \bibinfo{author}{\bibfnamefont{A.}~\bibnamefont{Fert}},
  \bibinfo{author}{\bibfnamefont{F.}~\bibnamefont{NguyenVanDau}},
  \bibinfo{author}{\bibfnamefont{F.}~\bibnamefont{Petroff}},
  \bibinfo{author}{\bibfnamefont{P.}~\bibnamefont{Eitenne}},
  \bibinfo{author}{\bibfnamefont{G.}~\bibnamefont{Creuzet}},
  \bibinfo{author}{\bibfnamefont{A.}~\bibnamefont{Friederich}},
  \bibnamefont{and} \bibinfo{author}{\bibfnamefont{J.}~\bibnamefont{Chazelas}},
  \bibinfo{journal}{Phys. Rev. Lett.} \textbf{\bibinfo{volume}{61}},
  \bibinfo{pages}{2472} (\bibinfo{year}{1988}).

\bibitem[{\citenamefont{Binasch et~al.}(1989)\citenamefont{Binasch,
  Gr{\"u}nberg, Saurenbach, and Zinn}}]{Binasch1989:PRB}
\bibinfo{author}{\bibfnamefont{G.}~\bibnamefont{Binasch}},
  \bibinfo{author}{\bibfnamefont{P.}~\bibnamefont{Gr{\"u}nberg}},
  \bibinfo{author}{\bibfnamefont{F.}~\bibnamefont{Saurenbach}},
  \bibnamefont{and} \bibinfo{author}{\bibfnamefont{W.}~\bibnamefont{Zinn}},
  \bibinfo{journal}{Phys. Rev. B} \textbf{\bibinfo{volume}{39}},
  \bibinfo{pages}{4828} (\bibinfo{year}{1989}).

\bibitem[{\citenamefont{Valet and Fert}(1993)}]{Valet1993:PRB}
\bibinfo{author}{\bibfnamefont{T.}~\bibnamefont{Valet}} \bibnamefont{and}
  \bibinfo{author}{\bibfnamefont{A.}~\bibnamefont{Fert}},
  \bibinfo{journal}{Phys. Rev. B} \textbf{\bibinfo{volume}{48}},
  \bibinfo{pages}{7099} (\bibinfo{year}{1993}).

\bibitem[{\citenamefont{Hershfield and Zhao}(1997)}]{Hershfield1997:PRB}
\bibinfo{author}{\bibfnamefont{S.}~\bibnamefont{Hershfield}} \bibnamefont{and}
  \bibinfo{author}{\bibfnamefont{H.~L.} \bibnamefont{Zhao}},
  \bibinfo{journal}{Phys. Rev. B} \textbf{\bibinfo{volume}{56}},
  \bibinfo{pages}{3296} (\bibinfo{year}{1997}).

\bibitem[{\citenamefont{{\v{Z}uti\'{c}}
  et~al.}(2001)\citenamefont{{\v{Z}uti\'{c}}, Fabian, and {Das
  Sarma}}}]{Zutic2001:PRB}
\bibinfo{author}{\bibfnamefont{I.}~\bibnamefont{{\v{Z}uti\'{c}}}},
  \bibinfo{author}{\bibfnamefont{J.}~\bibnamefont{Fabian}}, \bibnamefont{and}
  \bibinfo{author}{\bibfnamefont{S.}~\bibnamefont{{Das Sarma}}},
  \bibinfo{journal}{Phys. Rev. B} \textbf{\bibinfo{volume}{64}},
  \bibinfo{pages}{121201} (\bibinfo{year}{2001}).

\bibitem[{\citenamefont{Fabian et~al.}(2002)\citenamefont{Fabian,
  {\v{Z}uti\'{c}}, and {Das Sarma}}}]{Fabian2002:PRB}
\bibinfo{author}{\bibfnamefont{J.}~\bibnamefont{Fabian}},
  \bibinfo{author}{\bibfnamefont{I.}~\bibnamefont{{\v{Z}uti\'{c}}}},
  \bibnamefont{and} \bibinfo{author}{\bibfnamefont{S.}~\bibnamefont{{Das
  Sarma}}}, \bibinfo{journal}{Phys. Rev. B} \textbf{\bibinfo{volume}{66}},
  \bibinfo{pages}{165301} (\bibinfo{year}{2002}).

\bibitem[{\citenamefont{Bass and {Pratt Jr}}(2007)}]{Bass2007:JPCM}
\bibinfo{author}{\bibfnamefont{J.}~\bibnamefont{Bass}} \bibnamefont{and}
  \bibinfo{author}{\bibfnamefont{W.~P.} \bibnamefont{{Pratt Jr}}},
  \bibinfo{journal}{J. Phys.: Condens. Matter} \textbf{\bibinfo{volume}{19}},
  \bibinfo{pages}{183201} (\bibinfo{year}{2007}).

\bibitem[{\citenamefont{Rashba}(2002{\natexlab{b}})}]{Rashba2002:APL}
\bibinfo{author}{\bibfnamefont{E.~I.} \bibnamefont{Rashba}},
  \bibinfo{journal}{Appl. Phys. Lett.} \textbf{\bibinfo{volume}{80}},
  \bibinfo{pages}{2329} (\bibinfo{year}{2002}{\natexlab{b}}).

\bibitem[{\citenamefont{Jedema et~al.}(2001)\citenamefont{Jedema, Filip, and
  van Wees}}]{Jedema2001:Nature}
\bibinfo{author}{\bibfnamefont{F.~J.} \bibnamefont{Jedema}},
  \bibinfo{author}{\bibfnamefont{A.~T.} \bibnamefont{Filip}}, \bibnamefont{and}
  \bibinfo{author}{\bibfnamefont{B.~J.} \bibnamefont{van Wees}},
  \bibinfo{journal}{Nature} \textbf{\bibinfo{volume}{410}},
  \bibinfo{pages}{345} (\bibinfo{year}{2001}).

\bibitem[{\citenamefont{Huang et~al.}(2007)\citenamefont{Huang, Monsma, and
  Appelbaum}}]{Huang2007:PRL}
\bibinfo{author}{\bibfnamefont{B.}~\bibnamefont{Huang}},
  \bibinfo{author}{\bibfnamefont{D.~J.} \bibnamefont{Monsma}},
  \bibnamefont{and}
  \bibinfo{author}{\bibfnamefont{I.}~\bibnamefont{Appelbaum}},
  \bibinfo{journal}{Phys. Rev. Lett.} \textbf{\bibinfo{volume}{99}},
  \bibinfo{pages}{177209} (\bibinfo{year}{2007}).

\bibitem[{\citenamefont{Lubzens and Schultz}(1976)}]{Lubzens1976:PRL}
\bibinfo{author}{\bibfnamefont{D.}~\bibnamefont{Lubzens}} \bibnamefont{and}
  \bibinfo{author}{\bibfnamefont{S.}~\bibnamefont{Schultz}},
  \bibinfo{journal}{Phys. Rev. Lett.} \textbf{\bibinfo{volume}{36}},
  \bibinfo{pages}{1104} (\bibinfo{year}{1976}).

\bibitem[{\citenamefont{Jedema et~al.}(2003)\citenamefont{Jedema, Nijboer,
  Filip, and van Wees}}]{Jedema2003:PRB}
\bibinfo{author}{\bibfnamefont{F.~J.} \bibnamefont{Jedema}},
  \bibinfo{author}{\bibfnamefont{M.~S.} \bibnamefont{Nijboer}},
  \bibinfo{author}{\bibfnamefont{A.~T.} \bibnamefont{Filip}}, \bibnamefont{and}
  \bibinfo{author}{\bibfnamefont{B.~J.} \bibnamefont{van Wees}},
  \bibinfo{journal}{Phys. Rev. B} \textbf{\bibinfo{volume}{67}},
  \bibinfo{pages}{085319} (\bibinfo{year}{2003}).

\bibitem[{\citenamefont{Monod and {J\'{a}nossy}}(1977)}]{Monod1977:JLTP}
\bibinfo{author}{\bibfnamefont{P.}~\bibnamefont{Monod}} \bibnamefont{and}
  \bibinfo{author}{\bibfnamefont{A.}~\bibnamefont{{J\'{a}nossy}}},
  \bibinfo{journal}{J. Low Temp. Phys.} \textbf{\bibinfo{volume}{26}},
  \bibinfo{pages}{311} (\bibinfo{year}{1977}).

\bibitem[{\citenamefont{Johnson}(1993)}]{Johnson1993:PRL}
\bibinfo{author}{\bibfnamefont{M.}~\bibnamefont{Johnson}},
  \bibinfo{journal}{Phys. Rev. Lett.} \textbf{\bibinfo{volume}{70}},
  \bibinfo{pages}{2142} (\bibinfo{year}{1993}).

\bibitem[{\citenamefont{Mosendz et~al.}(2009)\citenamefont{Mosendz,
  Woltersdorf, Kardasz, Heinrich, and Back}}]{Mosendz2009:PRB}
\bibinfo{author}{\bibfnamefont{O.}~\bibnamefont{Mosendz}},
  \bibinfo{author}{\bibfnamefont{G.}~\bibnamefont{Woltersdorf}},
  \bibinfo{author}{\bibfnamefont{B.}~\bibnamefont{Kardasz}},
  \bibinfo{author}{\bibfnamefont{B.}~\bibnamefont{Heinrich}}, \bibnamefont{and}
  \bibinfo{author}{\bibfnamefont{C.~H.} \bibnamefont{Back}},
  \bibinfo{journal}{Phys. Rev. B} \textbf{\bibinfo{volume}{79}},
  \bibinfo{pages}{224412} (\bibinfo{year}{2009}).

\bibitem[{\citenamefont{Fabian and {\v{Z}uti\'{c}}}(2009)}]{Fabian2009:Julich}
\bibinfo{author}{\bibfnamefont{J.}~\bibnamefont{Fabian}} \bibnamefont{and}
  \bibinfo{author}{\bibfnamefont{I.}~\bibnamefont{{\v{Z}uti\'{c}}}}, in
  \emph{\bibinfo{booktitle}{From GMR to Quantum Information}}, edited by
  \bibinfo{editor}{\bibfnamefont{S.~B.} \bibnamefont{et~al.}}
  (\bibinfo{publisher}{Forschungszentrums {J{\"{u}}lich}},
  \bibinfo{year}{2009}), p.~\bibinfo{pages}{C1}.

\bibitem[{\citenamefont{Takahashi and Maekawa}(2003)}]{Takahashi2003:PRB}
\bibinfo{author}{\bibfnamefont{S.}~\bibnamefont{Takahashi}} \bibnamefont{and}
  \bibinfo{author}{\bibfnamefont{S.}~\bibnamefont{Maekawa}},
  \bibinfo{journal}{Phys. Rev. B} \textbf{\bibinfo{volume}{67}},
  \bibinfo{pages}{052409} (\bibinfo{year}{2003}).

\bibitem[{\citenamefont{Kochan et~al.}()\citenamefont{Kochan, Gmitra, and
  Fabian}}]{unpublished}
\bibinfo{author}{\bibfnamefont{D.}~\bibnamefont{Kochan}},
  \bibinfo{author}{\bibfnamefont{M.}~\bibnamefont{Gmitra}}, \bibnamefont{and}
  \bibinfo{author}{\bibfnamefont{J.}~\bibnamefont{Fabian}},
  \emph{\bibinfo{title}{(unpublished)}}.

\end{thebibliography}

\end{document}